\documentstyle[12pt,rotate,epsfig]{article}
\def\ngspace{\kern -.04em}

\catcode`@=11
\def\eqalign#1{\null\,\vcenter{\openup\jot\m@th
  \ialign{\strut\hfil$\displaystyle{##}$&$\displaystyle{{}##}$\hfil
      \crcr#1\crcr}}\,}
\catcode`@=12
\def\bln#1#2\eln{\begin{equation}\label{#1}%
   \eqalign{#2}\end{equation}\vskip3mm\noindent}

\def\brlist{\begin{thebibliography}{99}}
\def\brf{\bibitem}
\def\erlist{\end{thebibliography}}
\def\mcite{$\,$\cite}

\def\brlist{\begin{thebibliography}{99}}
\def\brf{\bibitem}
\def\erlist{\end{thebibliography}}
\begin{document}
\setlength{\baselineskip} {2.5ex}
\hyphenation {che-ren-kov}
\hyphenation {non-per-tur-ba-tive}
\hyphenation {per-tur-ba-tive}
\begin {center} 
{\bf Invited Talk at the\\ APS DNP Town Meeting 
on Electromagnetic \& Hadronic Physics\\
Dec. 2000, Jefferson Lab.\\
Tel Aviv U. Preprint TAUP 2661-2000}
\end{center}
\vspace{0.2cm}
\begin{center}
{\bf  Pion Polarizabilities and Hybrid Meson Structure\\ at 
CERN COMPASS
\\}
\vspace{0.2cm}
\end {center}
\vspace{0.2cm}
\begin {center}
{\bf
Murray Moinester\\
R. and B. Sackler Faculty of Exact
Sciences,\\
School of Physics,
Tel Aviv University, 69978 Ramat Aviv, Israel\\ 
E-mail: murraym@tauphy.tau.ac.il}
\end {center}
\vspace{1cm}
{\bf Abstract:\\}

\noindent

CERN COMPASS \cite {compass} can investigate pion-photon interactions, to
achieve a unique Primakoff Coulomb physics program, centered on pion
polarizability and hybrid meson structure studies \cite
{cd2,hadron1,bormio,hybrid}. COMPASS uses 100-280 GeV beams ($\mu$, $\pi$) 
and a virtual photon target, and magnetic spectrometers and calorimeters
to measure the complete kinematics of pion-photon reactions.  The COMPASS
experiment is scheduled to begin data runs in 2001.  Pion polarizabilities
and hybrid mesons can be studied via the Primakoff reactions $\pi^- \gamma
\rightarrow {\pi^-}' \gamma$ and $\pi^- \gamma \rightarrow Hybrid$.  The
electric $(\bar{\alpha})$ and magnetic $(\bar{\beta})$ pion and Kaon
polarizabilities characterize their deformation in an electromagnetic
field, as occurs during $\gamma\pi$ Compton scattering.  They depend on
the rigidity of their internal structures as composite particles, and are
therefore important quantities to test the validity of theoretical models.
The polarizability measurement will provide an important new test of QCD
chiral dynamics. The studies of quark-antiquark-gluon hybrid mesons would
improve our understanding of these exotic mesons.  COMPASS may improve
previous Primakoff polarizability and Hybrid studies by two to three
orders of magnitude. 

Appendixes A (Pion and Kaon Polarizabilities at COMPASS) and B (Hybrid
Meson Structure at COMPASS) of this contribution include evaluations
submitted to the APS DNP Town Meeting White Paper Committee.  These
summarize (1) the fundamental scientific issues addressed, (2) major
achievements since the last DNP long range plan, (3) the short and long
term U.S. outlook, (4) comparison of U.S. and global effort, (5) other
issues. 
 
\begin{center}
{\bf  1. Pion-Photon Interactions:\\}
\end{center}

\noindent

Pion polarizabilities and hybrid meson structure can be studied via
pion-photon interactions. Appendixes A and B of this contribution include
global evaluations on these subjects submitted to the APS DNP Town Meeting
White Paper Committee. The scientific background for these subjects is
described below. 

\vspace{.5cm}
\noindent
{\bf Pion Polarizabilities\\}

\noindent

\begin{figure}[tbc]
\centerline{\epsfig{file=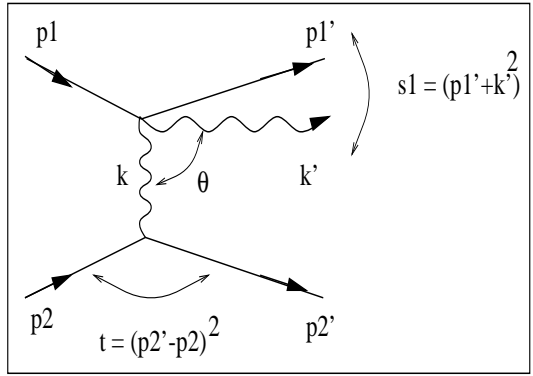,width=7cm,height=5cm}}
\caption{The Primakoff $\gamma$-hadron Compton process and kinematic variables
(4-momenta): p1, p1$^\prime$ = for initial/final hadron, p2, p2$^\prime$ = for 
initial/final target, k, k$^\prime$ = for initial/final gamma, and $\theta$ 
the scattering angle of the $\gamma$ in the alab frame.}
\label{fig:diagram1}
\end{figure}

For pion polarizability, $\gamma\pi$ scattering was measured (with large
uncertainties)  with 40 GeV pions \cite{anti1} via radiative pion
scattering (pion Bremsstrahlung) in the nuclear Coulomb field:  $\pi + Z
\rightarrow \pi' + \gamma + Z'.$ In the planned COMPASS, version of this
experiment, pion polarizability events have a stiff pion at angles smaller
than 0.5 mrad, very close to the non-interacting beam, and a single
forward photon at angles smaller than 2 mrad. The kinematic variables for
the pion polarizability Primakoff process are shown in
Fig.~\ref{fig:diagram1}. A virtual photon from the Coulomb field of the
target nucleus is scattered from the pion and emerges as a real photon
accompanying the pion at small forward angles in the laboratory frame,
while the target nucleus (in the ground state)  recoils with a small
transverse kick p$_t$. The peak at small target $p_t$ used to identify the
Primakoff process is precisely measured off-line using the beam and vertex
detectors. The radiative pion scattering reaction is equivalent to
$\gamma$ + $\pi^{-}$ $\rightarrow$ $\gamma$ + $\pi^{-}$ scattering for
laboratory $\gamma$'s of order 1 GeV incident on a target $\pi^{-}$ at
rest. It is an example of the well tested Primakoff formalism
\cite{jens,ziel2} that relates processes involving real photon
interactions to production cross sections involving the exchange of
virtual photons. 

For the $\gamma\pi$ interaction at low energy, the $\chi$PT effective
Lagrangian establishes relationships between different processes. For
example, using data from radiative pion beta decay, $\chi$PT predicts the
pion polarizabilities \cite{gass1,hols1}:  $\bar{\alpha}_{\pi}$ =
-$\bar{\beta}_{\pi}$ = 2.7 $\pm$ 0.4, expressed in units of $10^{-43}$
cm$^3$.

The pion polarizabilities deduced by Antipov et al. \cite{anti1} in their
low statistics experiment ($\sim$ 7000 events) were $\bar{\alpha}_{\pi} =
-\bar{\beta}_{\pi} = 6.8 \pm 1.4 \pm 1.2$. It was assumed in the analysis
that $\bar{\alpha}_{\pi} + \bar{\beta}_{\pi} = 0$, as expected
theoretically \cite {hols1}.  The deduced polarizability value, ignoring
the large error bars, is about three times larger than the $\chi$PT
prediction. {\bf The available polarizability results \cite {babu2} have
large uncertainties. There is a clear need for the new and improved
radiative pion scattering data from COMPASS.}

  New COMPASS data will be compared also to new Mainz data.  At MAMI-B at
Mainz, measurements \cite {mainz} and calculations \cite {un00} are under
way of $p(\gamma,n \pi^+ \gamma')$ radiative pion photoproduction reaction
on the proton. The elastic $\gamma \pi^+$ scattering cross section can be
found by extrapolating such data to the pion pole. This corresponds to
"Compton" scattering of gamma's from virtual $\pi^+$ targets in the
proton, and therefore also allows a measure of the pion polarizability.
The experiment is running with 500-800 MeV tagged photons, with detection
of $\gamma'$, neutron, and charged pion in coincidence. For the long term,
a pion polarizability experiment with polarized tagged photons was
proposed \cite {hi99}, associated with the proposed JLab 12 GeV upgrade.

\vspace{.5cm}
\noindent
{\bf Hybrid Mesons\\}

\noindent

The hybrid ($q\bar{q}g$) mesons, along with glueballs ($gg$) are some of
the most interesting consequences of the non-Abelian nature of QCD. The
unambiguous confirmation of hybrid states will be a major event in hadron
spectroscopy.  Hybrids contain explicit glue as opposed to hidden glue in
conventional hadrons.  Understanding explicit glue is critical,
considering that most of the mass around us is made of gluons.  For the
understanding of confinement, it is of major importance and relevance to
establish the existence of hybrid mesons and to study their structure. 
Input from experiments is needed to guide us to better understanding. 
Evidence from completely different experiments are needed to show that the
present evidence is not the result of some incorrectly interpreted
artifice. We may look forward to comparisons of new COMPASS and JLab \cite
{jlabhybrid,azas,HallD} hybrid meson experiments.

Detection of these exotic states is a long-standing experimental puzzle. 
The most popular approach for hybrid searches is to look for the
`oddballs'---mesons with quantum numbers not allowed for ordinary
$q\bar{q}$ states.  For Primakoff/diffractive production, the `oddball'
mesons for $J\leq 2$ are: $I^G(J^{PC})=1^-(1^{-+})$ `$\pi_1$'---or more
generally $I^G(J^{PC})=1^+(0^{+-})$ `$b_0$' or $I^G(J^{PC})=1^+(2^{+-})$
`$b_2$'---hybrids. The signature for such a state is when a detailed
partial wave analysis (PWA) of a large data sample requires a set of
quantum numbers which is inconsistent with a regular (q-qbar) meson.

Barnes and Isgur, using the flux-tube model \cite {tb,ni}, calculated the
mass of the lightest gluonic hybrid to be around 1.9 GeV, with the quantum
numbers of $J^{PC}=1^{-+}$.  Close and Page \mcite{fluxtb} predict that
such a gluonic hybrid should decay into the following channels: 

\begin{center}\begin{tabular}{c|c|c|c|c}
   $b_1\pi$&$f_1\pi$&$\rho\pi$&$\eta\pi$&$\eta'\pi$\\
\hline
170&60&$5\to 20$&$0\to 10$&$0\to 10$\\
\end{tabular}\end{center}

\noindent
where the numbers refer to the partial widths in MeV.  They expect its
total width to be larger than 235-270 MeV, since the $s+\bar s$ decay
modes were not included.  Recent updates on hybrid meson structure are
given in Refs. \cite{nib98,kpb98,sgjn}

   From more than a decade of experimental efforts at IHEP \cite
{ihep1,ihep2,ves}, CERN \cite {na12,cb}, KEK \cite{kek}, and BNL \cite
{E852}, several hybrid candidates have been identified. More recently, BNL
E852 \cite {E852} reported two $J^{PC}= 1^{-+}$ resonant signals at masses
of 1.4 and 1.6 GeV in $\eta\pi^-$ and $\eta\pi^0$ systems, as well as in
$\pi^+ \pi^- \pi^-$, $\pi^- \pi^0 \pi^0$, $\eta' \pi^-$ and $f_1(1285) 
\pi^-$.  The VES collaboration presented \cite {whs} the results of a
coupled-channel analysis of the $\pi_1(1600)$ meson in the channels
$\rho\pi$, $\eta'\pi$ and $b_1(1235)\pi$. Their results, 
and those of the crystal barell collaboration \cite {cb}, are consistent
with the BNL results. 

\begin{figure}[tbc]
\centerline{\epsfig{file=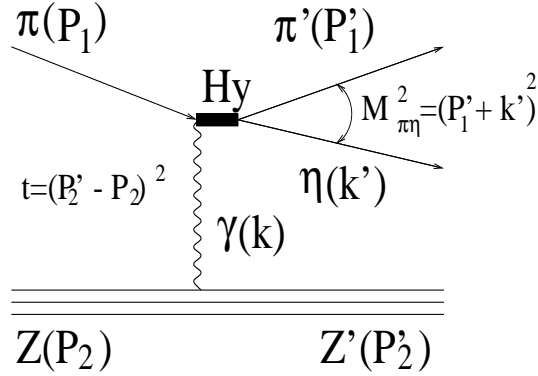,width=7cm,height=5cm}}
\caption{The Primakoff $\gamma$-pion Hybrid production process and
kinematic variables (4-momenta): P1, P1$^\prime$ = for initial/final pion,
P2, P2$^\prime$ = for initial/final target, k = for initial $\gamma$,
k$^\prime$ = for final $\eta$. By VDM, the exchanged $\gamma$ behaves
like a $\rho$.}
\label{fig:diagram}
\end{figure}

The kinematic variables for the $\pi \gamma \rightarrow HY \rightarrow
\pi^- \eta$ Primakoff process in COMPASS are shown in
Fig.~\ref{fig:diagram}. A virtual photon from the Coulomb field of the
target nucleus interacts with the pion, a Hybrid meson is produced and
decays to $\pi^- \eta$ at small forward angles in the laboratory frame,
while the target nucleus (in the ground state)  recoils coherently with a
small transverse kick p$_T$. The peak at small target $p_T$ used to
identify the Primakoff process is measured off-line using the beam and
vertex detectors.

   The partial-wave analysis (PWA) of systems such as $\eta\pi$ or $\eta'
\pi$ in the mass region below 2 GeV requires care and experience. This is
so because (1) this region is dominated by the strong $2^+$ `background'
($a_2$ resonance), and (2) that the PWA may give ambiguous results \cite
{ihep2} for the weaker $1^{-+}$ wave. For Primakoff production, the hybrid
production cross section may increase relative to the close lying $a_2$
state, considering the estimated radiative widths. These are $\Gamma(a_2
\rightarrow \pi \gamma)=300$~ke$\!$V, and $\Gamma(\pi_1 \rightarrow \pi
\gamma)\approx 90-540$~ke$\!$V, as discussed in Section 3.  This may
significantly diminish the partial wave analysis uncertainties of
non-Primakoff production experiments for the $1^{-+}$ wave. Furthermore,
in Primakoff (photon exchange) and diffractive (glue-rich pomeron
exchange)  hybrid production, meson exchange backgrounds and final state
interactions are strongly suppressed, which is an important advantage
compared to previous $\pi^-p \rightarrow Hybrid$ production experiments. 

   For both BNL E852 and VES data, it is not known what Regge exchanges
are responsible for the production of the $J^{PC}=1^{-+}$ exotic states at
1.4 and 1.6 GeV, Both the $a_2(1320)$ and the exotic waves are produced
via natural-parity exchanges which include the Pomeron.  If Pomeron
exchange is indeed responsible for the production, then diffractive
production in COMPASS can provide an additional handle with which to
tackle the study of exotic waves. 

   One can succinctly summarize the situation as follows: a production of
the wave $I^G(J^{PC})=1^-(1^{-+})$ is dependent on the strength of the
$\pi\rho$ decay modes in the case of the Primakoff production.  And both
BNL and VES claim this decay mode for the different Hybrid candidates. 
For diffractive production, the relative strengths depend on the supposed
decay modes $\pi_1(1400)\pi$ and $\pi_1(1600)\pi$ of the tensor glueball
(2$^{++}$), since the Pomeron is thought to be on the Regge trajectory
corresponding to the tensor glueball with a presumed mass around 2 GeV.
Corresponding to the glueball decay $G(2^{++}) \rightarrow \pi^+ Hybrid$,
one expects diffractive production via $\pi^- G(2^{++})  \rightarrow
Hybrid$.  This is an additional strong advantage of the COMPASS hybrid
meson study. We can look forward to two complementary production modes of
exotic mesons, increasing our chance for achieving a decisive advance on
our understanding of the meson constituents. COMPASS can study Primakoff
and diffractive production of non-strange light-quark hybrid mesons in the
1.4-2.5 GeV mass region, including all the hybrid candidates from previous
studies. 

\noindent
\begin{center}
{\bf 2.~Experimental Requirements\\}
\end{center}

We considered \cite {cd2,hadron1,bormio,hybrid} the beam, target,
detector, and trigger requirements for polarizability and hybrid meson
studies, with minimum background contamination. Here we discuss briefly
only the electromagnetic calorimeter and the Primakoff trigger. 

\vspace{.5cm}
\noindent
{\bf The COMPASS Electromagnetic Calorimeter\\}

COMPASS has a 2 meter diameter EM calorimeter, which is so far
instrumented for the central 1 meter diameter.  Funding for ADC readout
electronics is still needed to be able to utilize the full 2 meter
diameter coverage.  As can be seen in Fig.~\ref{fig:diagram}, COMPASS
needs to also detect $\eta$s for the hybrid study.  The two $\gamma$s from
$\eta$ decay have half-opening angles $\theta_{\gamma\gamma}^h$ for the
symmetric decays of $\theta_{\gamma\gamma}^h= m/E_{\eta}$, where m is the
mass ($\eta$)  and E$_{\eta}$ is the $\eta$ energy. Opening angles are
somewhat larger for asymmetric decays. In order to catch about 50\% of the
decays, it is necessary to subtend a cone with double that angle, i.e. 
$\pm 2m/E_{\eta}$, neglecting the angular spread of the original $\eta$s
around the beam direction.  Consider an ECAL2 $\gamma$ detector with a
circular active area with 2 m diameter.  Consider the $\pi\eta$ channel. 
For an ECAL2 of 1 m radius at 30 m from the target, $\eta$s above
E$_{\eta}$=33 GeV are therefore accepted.  At half this energy, the
acceptance practically vanishes. The acceptance of course depends on the
Hybrid mass, mostly between 1.4 and 2.5 GeV for the planned COMPASS study.
Detailed Monte Carlo studies are in progress for the different possibly
Hybrid decay modes, for a range of assumed masses.  For the $\pi$f$_1$
channel, with for example $f_1 \rightarrow \pi\pi\eta$, the $\eta$s will
have low energy, and therefore large gamma angles. To maintain good
acceptance for low energy $\eta$s, the ECAL2 diameter should be 2 m or
more. 

\vspace{.5cm}
\noindent
{\bf The COMPASS Primakoff Trigger\\}

\begin{figure}[tbc] 
\centerline{\epsfig{file=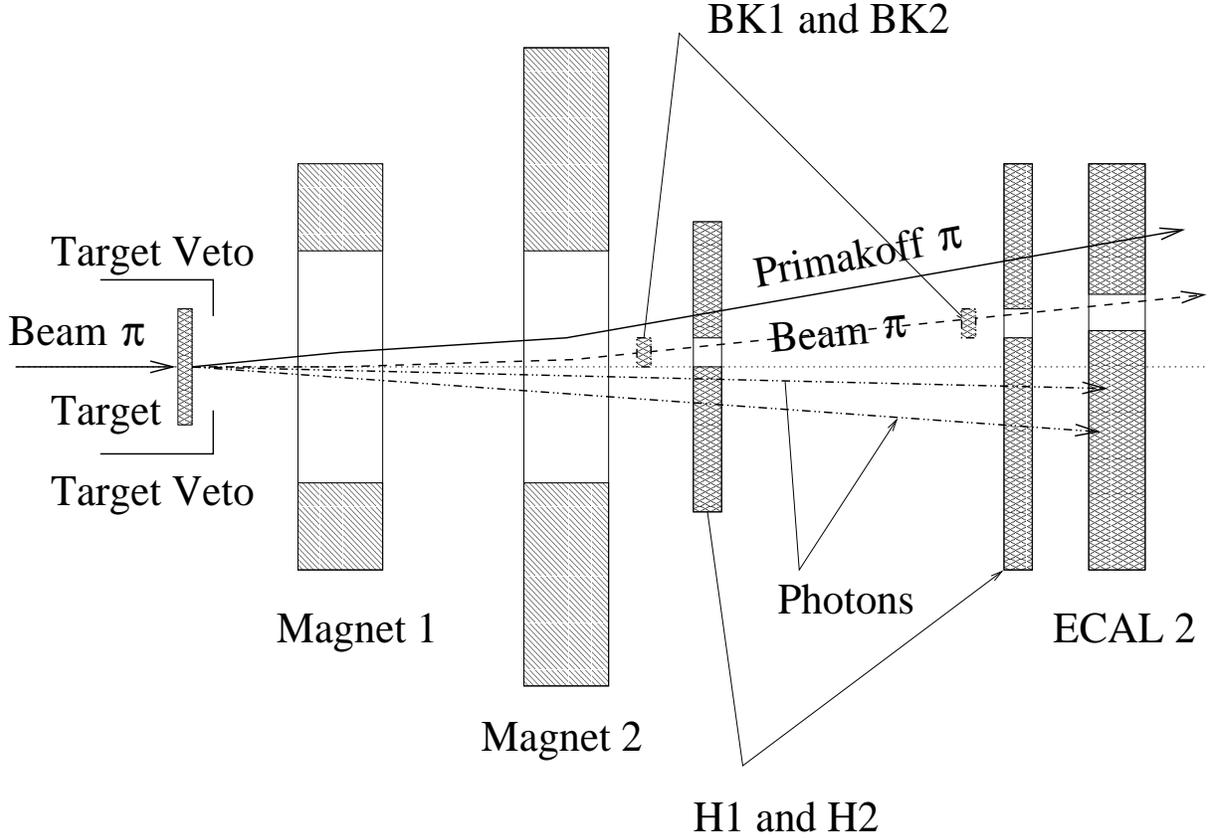,width=16cm,height=11cm}}
\caption{Detector layout for the COMPASS Primakoff Hybrid trigger.
BK1,BK2=beam killer system, H1,H2=hodoscope system for charged particle 
detection, ECAL2=second photon calorimeter.}
\label{fig:trigger}
\end{figure}

We design \cite {cd2,hadron1,bormio,hybrid} the COMPASS
Primakoff/Diffractive hybrid meson trigger to enhance the acceptance and
statistics, and also to yield a trigger rate closer to the natural rate
given by the Compton scattering and hybrid cross sections.  We may veto
target break-up events via veto scintillators around the target. For
polarizability and hybrid meson physics, the trigger uses the
characteristic decay pattern: one or three charged mesons with gamma hits,
or three charged mesons and no gamma hits.  The trigger \cite
{cd2,hadron1,bormio,hybrid} for the $\pi\eta$ hybrid decay and
polarizability channel (charged particle multiplicity =1) is based on a
determination of the pion energy loss (via its characteristic angular
deflection), correlated in downstream scintillator hodoscopes stations (H1
versus H2)  with the aid of a fast matrix chip, as shown in
Fig.~\ref{fig:trigger}.  We also test alternative and/or complementary
trigger concepts.  For example, the non-interacting beam may be detected
and vetoed by the Beam Kill veto trigger detectors BK1/BK2, which follow
the pion trajectory, as shown in Fig. Fig.~\ref{fig:trigger}.
 
We studied \cite {cd2,hadron1,bormio} the acceptance for this trigger
using the MC code POLARIS, which generates Primakoff pion-photon
(polarizability) interactions, with realistic beam phase space. The
simulation was done for beam momentum of 190 GeV/c.  Using the beam killer
system maintains good acceptance for Primakoff pions of momenta $<$ 160
GeV/c and photons of momenta $>$ 30 GeV/c.  Introducing the lower
threshold for the ECAL2 signal to be equal 20 GeV, helps to suppress
background processes, but does not affect acceptance in its most efficient
region.  Finally, for the given trigger design, we achieve a large and
flat acceptance versus photon Compton scattering angles \cite {bormio}. 
This is important to extract reliably the pion polarizability by a fit of
the data to the theoretical cross section.  This trigger does not affect
the acceptance at the important back-angles where the polarizability
contribution is largest. 

\begin{center}
{\bf  3. Objectives and Expected Significance\\}
\end{center}

\noindent

We studied the statistics attainable and uncertainties achievable for the
pion polarizabilities in the COMPASS experiment, based on Monte Carlo
simulations. We begin with an estimated 0.5 mb Compton scattering cross
section per Pb nucleus and a total inelastic cross section per Pb nucleus
of 0.8 barn.  High statistics will allow systematic studies, with fits
carried out for different regions of photon energy $\omega$, Z$^2$, etc.; 
and polarizability determinations with statistical uncertainties of order
0.2. For the kaon polarizability, due to the lower beam intensity, the
statistics will be roughly 50 times lower. A precision kaon polarizability
measurement requires more data taking time.

For pion polarizability, in four months of running, we obtain 1.4 $\times$
10$^{13}$ beam pions. With a 1~\% interaction length target, we obtain 1.4
$\times$ 10$^{11}$ interactions. Based on the cross sections above, the
Primakoff event rate R (events per interaction) is R=6.3 $\times$
10$^{-4}$. Therefore, in a 4 month run, one obtains 8.8 $\times$ 10$^{7}$
Primakoff polarizability events at 100\% efficiency.  Considering
efficiencies for tracking, $\gamma$ detection, accelerator operation,
trigger, we estimate a global efficiency of $\epsilon$(total)=0.11, or
9.6$\times$ 10$^6$ useful events per 4 month run.  This is roughly a
factor 1000 higher than the previous polarizability experiment.

We make rough estimates of the statistics attainable for hybrid production
in the COMPASS experiment. Monte Carlo simulations in progress will refine
these estimates. We assume a 125-1250$\mu$b Hybrid meson production cross
section per Pb nucleus (near 1.5 GeV mass). This estimate is based on two
considerations. First, a straightforward application of VDM with
$\rho-\gamma$ coupling g$_{\rho\gamma}^2/\pi$=2.5, gives a width of
$\Gamma(\pi_1 \rightarrow \pi \gamma$) = 75-750 keV for a 1.5 GeV Hybrid,
assuming $\Gamma(\pi_1 \rightarrow \pi \rho$)= 10-100 MeV, a range
corresponding to 3.3-33\% of the claimed 1.5 GeV hybrid width. Integrating
the Primakoff Hybrid production differential cross section for a 280 GeV
pion beam with this $\Gamma(\pi_1 \rightarrow \pi \gamma$) width gives
125-1250 $\mu$b. Second, a FNAL E272 measurement indicated (but with high
uncertainty) that $\Gamma(\pi_1\rightarrow\pi\gamma) \times
BR(\pi_1\rightarrow\pi f_1)  \approx 250$ keV for a 1.6 GeV Hybrid
candidate. This would be consistent with the above maximum VDM
$\Gamma(\pi_1 \rightarrow \pi \gamma$) estimate for BR($\pi_1 \rightarrow
\pi f_1$)  = 33\%.  With a total $\pi$ inelastic cross section per Pb
nucleus of 0.8 barn, the Primakoff Hybrid production event rate R (events
per interaction) is then R= 1.6-16 $\times$ 10$^{-4}$. 

In four months of running, we obtain 1.4 $\times$ 10$^{13}$ beam pions. 
With a 1\% interaction length target, we obtain 1.4 $\times$ 10$^{11}$
interactions. Therefore, one may obtain 2.2-22$\times$10$^{7}$ Hybrid
Primakoff events at 100\% efficiency.  We assume now a 50\% accelerator
operation efficiency. We also estimate a global 10\% average detection
efficiency over all decay channels for tracking, $\gamma$ detection,
$\eta$ acceptance and identification, trigger acceptance, global geometric
acceptance, and event reconstruction efficiency.  All these effects give a
global efficiency of 5\%. Therefore, we may expect to observe a total of
1.1-11$\times$10$^{6}$ Hybrid decays in all decay channels. For example,
following the Close and Page predictions, we may expect 24\% in $\pi f_1$,
2-8\% in $\pi \rho$, 67\% in b$_1\pi$, 0-4\% in $\eta\pi$, 0-4\% in
$\eta'\pi$, etc. 

For 2, 2.5, 3.0 GeV mass Hybrids, the number of useful events decreases by
factors of 6, 25, and 100, respectively. But even in these cases, assuming
again a global 5\% efficiency, that represents very interesting potential
samples of 1.8-18$\times$10$^{5}$, 4.4-44$\times$10$^{4}$, and
1.1-11$\times$10$^{4}$ Hybrid meson detected events, with masses 2, 2.5,
and 3 GeV respectively.

COMPASS can study hybrid meson candidates near 1.4, 1.6, 1.9 GeV produced
by the Primakoff and Diffractive processes. COMPASS should also be
sensitive to pionic hybrids in the 2-3 GeV mass range. We may obtain
superior statistics for hybrid states if they exist, and via a different
production mechanism, without possible complication by hadronic final
state interactions. We may also get important data on the different decay
modes for this state. The observation of these/other hybrids in different
decay modes and in a different experiment would constitute the next
important step following the evidence so far reported. 

COMPASS provides a unique opportunity to investigate pion polarizabilities
and QCD hybrid exotics. Taking into account the very high beam intensity,
fast data acquisition, high acceptance and good resolution of the COMPASS
setup, one can expect from COMPASS the highest statistics and a
`systematics-free' data sample that includes many tests to control
possible systematic errors. Intercomparisons between COMPASS and past plus
new experiments \cite {mainz,GSI,chpc}, with complementary methodologies,
should allow fast progress on understanding pion polarizabilities and
hybrid meson structure, and on fixing the systematic uncertainties.

\begin{center}
4. {Acknowledgments:\\}
\end{center}

This work was supported by the Israel Science Foundation founded by the
Israel Academy Sciences and Humanities, Jerusalem, Israel.  Thanks are due
to F. Bradamante, A. Bravar, D. Casey, S. U. Chung, M. Faessler, T. 
Ferbel, O. Gavrishuk, Y. Khokhlov, M. Lamanna, J. Lichtenstadt, G. Mallot,
A. Olchevski, S. Paul, E. Piasetzky, J. Pochodzalla, V. Poliakov, S. 
Prakhov, I. Savin, L. Schmitt H.-W. Siebert, V. Steiner, D. von Harrach,
and Th. Walcher for valuable discussions. 

\brlist

\bibitem{compass} F. Bradamante, S. Paul et al., CERN Proposal COMPASS,\\
http://wwwcompass.cern.ch/, CERN/SPSLC 96-14, SPSC/P 297.  \\

\brf{cd2} M. A. Moinester, V. Steiner, Pion and Kaon Polarizabilities and
Radiative Transitions, Proc. `Chiral Dynamics Workshop' 
U. Mainz, Sept. 1997, hep-ex/9801008. 

\bibitem{hadron1} M. A. Moinester and V. Steiner, Primakoff Physics for
CERN COMPASS, JINR (Dubna) CERN COMPASS Summer School, Prague, Czech
Republic, Aug. 1997, hep-ex/9801011. 

\bibitem{bormio} M. A. Moinester, V. Steiner, S. Prakhov, Hadron-Photon
Interactions in COMPASS, Proc. XXXVII Meeting on Nuclear Physics, Bormio,
Italy, Jan. 1999, hep-ph/9910039

\bibitem{hybrid} M. A. Moinester, S. U. Chung, Hybrid Meson Structure at
COMPASS, Contributions to COMPASS WEEK, Munich, Germany,
Oct. 1999, hep-ex/0003008. 

\bibitem {anti1}Yu. M. Antipov {\em et al.}, Phys. Lett. {\bf 121B}, 445
(1983).

\brf{jens} T. Jensen {\em et al.}, Phys. Rev. {\bf 27D}, 26 (1983).

\bibitem {ziel2} M. Zielinski {\em et al.}, Phys. Rev. {\bf 29D}, 2633
(1984).

\bibitem {gass1}  J.Gasser and H.Leutwyler, Nucl. Phys. {\bf B250}, 465
(1985).

\bibitem {hols1}   B. R. Holstein, Comments Nucl. Part. Phys. {\bf 19},
239 (1990).

\bibitem {babu2} D. Babusci, S. Bellucci, G. Giordano, G. Matone,
A. M. Sandorfi, M. A. Moinester,\\ Phys. Lett. {\bf B277}, 158 (1992).

\bibitem{mainz} J.~Ahrens {\em et al.}, Measurement of the $\pi$ Meson
Polarizability,\\ U. Mainz MAMI A2 Proposal, Few-Body Suppl. {\bf 9},
449 (1995).  

\bibitem {un00} C. Unkmeir, U. Mainz, Ph.D. thesis, 2000. 

\bibitem {hi99} R. Hicks {\em et al.}, JLab Proposal E99-010. 

\bibitem {jlabhybrid} N. Isgur, A. Dzierba, CERN Courier {\bf 40}, 
23 (2000);\\ http://www.cerncourier.com/main/article/40/7/16

\bibitem {azas} A. Dzierba {\em et al.}, American Scientist {\bf 88},
406 (2000);\\http://www.americanscientist.org/articles/00articles/dzierba.html
 
\bibitem {HallD} A. Dzierba {\em et al.}, The Hall D project Design
Report, Searching for QCD Exotics with a Beam of Photons, Nov. 2000;
http://dustbunny.physics.indiana.edu/HallD/

\bibitem {tb} T. Barnes, Photoproduction of Hybrid Mesons, Cont. to
Confinement III, Newport News, VA, June 1998, nucl-th/9907020

\bibitem {ni} N. Isgur, Phys. Rev. {\bf D60}, 114016 (1999). 

\bibitem {fluxtb} F. Close, P. Page, Nucl. Phys. {\bf B443}, 
233 (1995) 

\bibitem {nib98} N. Isgur, Spectroscopy - An Introduction and Overview,
Proc. of Baryons 98, Bonn, Germany, Sept.  1998, Eds. D. W. Menze, B.
Metsch, World Scientific, 1999. 

\bibitem {kpb98} K. Peters, Meson Spectroscopy and Exotic Quantum Numbers,
Proc. of Baryons 98, Bonn, Germany, Sept.  1998, Eds. D. W. Menze, B. 
Metsch, World Scientific, 1999.

\bibitem {sgjn} S. Godfrey, J. Napolitano, Rev. Mod. Phys. {\bf 71}, 1411
(1999). 

\brf{ihep1} D. Alde {\em et al.}, Proc. of HADRON-97, BNL, August 1997.

\brf{ihep2} Yu.D.Prokoshkin and S.A.Sadovsky, Phys. At. Nucl. {\bf 58},
606 (1995).

\brf{ves} G. M. Beliadze {\em et al.}, Phys. Lett. {\bf } B313 (1993) 
276;\\ A.Zaitsev, Proc. of HADRON-97, BNL. August 1997. 

\brf{na12} D. Alde et at., Phys. Lett. {\bf B205}, 397 (1988).

\brf{cb} A. Abele {\em et al.}, Phys. Lett. {\bf 423B}, 175 (1998).

\brf{kek} H. Aoyagi {\em et al.}, Phys. Lett. {\bf B314}, 246 (1993).

\brf{E852} D. R. Thompson {\em et al.}, Phys. Rev. Lett. {\bf 79}, 1630
(1997);  G. S. Adams {\em et al.}, Phys. Rev. Lett. {\bf 81}, 5760 (1998);
P.  Eigenio, hep-ph/0010337. 

\bibitem {whs} V. Dorofeev, New Results from VES, Workshop on Hadron
Spectroscopy, Rome, Italy, March 1999, hep-ex/9905002; Yu. Khokhlov {\em
et al.}, Nucl. Phys. {\bf A663}, 596 (2000). 

\bibitem {GSI} U. Wiedner {\em et al.}, Hybrid Meson prospects via a new
facility for $p \bar{p}$ reactions, www.gsi.de/GSI-Future/program.html,
Darmstadt, Germany, Oct.  2000. 

\bibitem {chpc} S. U. Chung, Hybrid Meson Prospects via CERN LEP
two-photon and FNAL CDF/D0 double-Pomeron production, private
communication. 

\erlist

\newpage
\baselineskip=18.0pt
\parskip=12.0pt
\parindent=2em
\noindent 
{\bf Appendix A, submitted to APS DNP White Paper Committee}
\section {PION AND KAON POLARIZABILITIES AT COMPASS}
\label{sec:hypol}

\noindent {MURRAY MOINESTER}
\vspace {0.2in}

\noindent {\bf The fundamental scientific questions addressed}\\

\noindent 

The electric $(\bar{\alpha})$ and magnetic $(\bar{\beta})$ pion and Kaon
polarizabilities characterize their deformation in an electromagnetic
field, as occurs during $\gamma\pi$ or $\gamma$-Kaon Compton scattering. 
They depend on the rigidity of their internal structures as composite
particles, and are therefore important quantities to test the validity of
theoretical models. Pion (Kaon) polarizabilities can be studied at CERN
COMPASS \cite {compass} via pion and Kaon Primakoff reaction such as
$\pi^- \gamma \rightarrow {\pi^-}' \gamma$ Compton scattering \cite
{cd2,hadron1,bormio}. In pion-photon Primakoff scattering, a high energy
pion scatters from a virtual photon in the Coulomb field of the target
nucleus.  The pion polarizabilities are determined by their effect on the
shape of the measured $\gamma \pi$ Compton scattering angular
distribution. For theory, the $\chi$PT effective Lagrangian \cite
{gass1,hols1}, using data from radiative pion beta decay, predicts the
pion electric and magnetic polarizabilities $\bar{\alpha}_{\pi}$ =
-$\bar{\beta}_{\pi}$ = 2.7 $\pm$ 0.4, expressed in units of $10^{-43}$
cm$^3$. For the kaon, the $\chi$PT polarizability prediction \cite {hols1}
is $\bar{\alpha}_{\pi}$ = 0.5. But available experimental pion
polarizability results \cite {anti1,babu2} cover a large range of values
and have large uncertainties. And Kaon polarizability measurements have
never been carried out. New high precision pion and Kaon polarizability
measurements will therefore provide important new tests of QCD chiral
dynamics.

\noindent {\bf Major achievements since the last Long Range Plan}\\

\noindent 

(1) For pion polarizabilities, a measurement is in progress
\cite {mainz} at MAMI-B at Mainz, via the $p(\gamma,n \pi^+ \gamma')$
radiative pion photoproduction reaction. The $\gamma \pi^+$ Compton
scattering cross section can be found by extrapolating such data to the
pion pole, and thereby allows a measure of the pion polarizability. 
Theoretical studies were carried out \cite {un00} and will be continued,
to minimize the errors associated with the extrapolation to the pion
pole.\\

(2) The COMPASS experiment was approved by CERN. The Y2K setup run
included a Primakoff test run with a 1 meter diameter EM calorimeter for
the pion polarizability study. During the coming years of data taking,
COMPASS run time will be shared between muon (gluon polarization) and pion
beam physics programs. During muon runs, setup and normalization of the
pion polarizability program may proceed via measurements of muon-photon
Primakoff scattering. \\ 

(3) A pion polarizability experiment with polarized tagged photons has
been proposed \cite {hi99} for the JLAB 12 GeV upgrade. 

\noindent {\bf Short and Long term ($<$3 yrs and long term $<$10 yrs) U.S. 
outlook}

For the long term, JLab may run and analyze its pion polarizability
experiment. 

\noindent {\bf Comparison of U.S. and global effort}

The U.S. and global efforts contine strongly on the experimental and
theoretical fronts for pion polarizabilities. The aim is to achieve
consistent results in different experiments. Only then will it be possible
to unambiguously establish definitive values for pion polarizabilities.
COMPASS plans to use a 200 GeV pion beam and a photon target, and magnetic
spectrometers and electromagnetic calorimeters, to measure pion-photon
Compton scattering, and the pion polarizability.  Primakoff scattering has
the advantages also that meson exchange backgrounds and final state
interactions are strongly suppressed. 

For the short term, analysis of the Mainz MAMI-B data run, scheduled for
completion Jan. 2001, should lead to a new pion polarizability result.
Also, the COMPASS experiment begins its muon and pion beam physics
programs in 2001. Associated Monte Carlo simulations are in progress.  For
the long term, COMPASS, JLab, and Mainz (MAMI-C A2 collaboration at higher
tagged photon energies)  may run and analyze and compare data from
different experiments for pion polarizabilities. Only COMPASS may measure
Kaon polarizabilities.  Such data and intercomparisons will be valuable
for fixing the systematic uncertainties in pion and Kaon polarizability
measurements. 

\noindent {\bf Other issues}

For the kaon polarizability program, a modest COMPASS upgrade will be
required to achieve pion/Kaon beam particle identification. One may
achieve a tagged Kaon beam intensity of order 1 MHz, by a photomultiplier
detector upgrade (to allow high beam intensities)  of the exisiting
COMPASS CEDARS Cerenkov detectors.

\newpage
\baselineskip=18.0pt
\parskip=12.0pt
\parindent=2em
\noindent {\bf Appendix B, submitted to APS DNP White Paper Committee} 
\section{HYBRID MESON STRUCTURE AT COMPASS}
\label{sec:hypol}
\noindent {MURRAY MOINESTER}
\vspace {0.2in}

\noindent {\bf The fundamental scientific questions addressed}\\

\noindent 

To understand confinement, it is of major importance and relevance to
establish the existence of hybrid mesons and to study their structure. The
hybrid ($q\bar{q}g$) mesons contain explicit glue as opposed to hidden
glue in conventional hadrons. Understanding explicit glue is critical,
considering that most of the mass around us is made of gluons.  COMPASS
\cite {compass} may study Primakoff and diffractive production of
non-strange light-quark hybrid mesons in the 1.4-3.0 GeV mass region,
including hybrid candidates from previous studies \cite {bormio,hybrid}. 

\noindent {\bf Major achievements since the last Long Range Plan}\\

\noindent 

(1) From the experimental efforts at BNL, IHEP, KEK, CERN, a number of
$J^{PC}= 1^{-+}$ resonant signals were reported \cite
{ihep1,ihep2,ves,na12,cb,kek,E852,whs} at masses between 1.4 and 1.9 GeV
in a variety of decay channels, including the $\rho\pi$ channel. 
Confidence in hybrid mesons is increasing, but further complementary
evidence is still sorely needed.\\

(2) Theoretical studies of hybrid mesons are shedding new light on their
structure. As examples, Barnes and Isgur \cite {tb,ni} via the flux-tube
model, recently calculated the mass of the lightest gluonic hybrid to be
around 1.9 GeV, with the quantum numbers of $J^{PC}=1^{-+}$;  while Close
and Page \cite {fluxtb} predict the decay modes of such a gluonic
hybrid.\\

(3) The COMPASS experiment was approved by CERN. It already had an
equipment setup run in summer 2000, including a test run with a 1 meter
diameter EM calorimeter.  During the coming years of data taking, run time
will be shared between muon and pion beam physics programs. Besides the
Primakoff program, COMPASS runs will involve studies of gluon polarization
in the proton, using the muon beam to study photon-gluon fusion. \\

(4) GSI Darmstadt \cite {GSI} plans a hybrid meson program via a new
facility for $p \bar{p}$ reactions. \\

(5) The planned 12 GeV JLab upgrade would allow hybrid meson studies at
JLab Hall D \cite {jlabhybrid,azas,HallD}.\\

(6)  Analysis of CERN LEP data \cite {chpc} may reveal two-photon
production of exotic mesons.  From the vector-dominance model, it is clear
that two-photon systems should be rich in 4-quark exotics. Similarly, the
Fermilab CDF (no Roman pots) and D0 (Roman pots being built) can give
double-Pomeron production of exotic mesons \cite {chpc}. Pomerons are
gluonic; the mesons produced could be rich in gluonic hybrids. 

\noindent {\bf Short and long term ($<$3 yrs and long term $<$10 yrs) U.S. 
outlook}\\

Further BNL data (from analysis of completed experiments) and further
theoretical calculations are becoming available.  For the long term, JLab
plans to take and analyze data for hybrid meson structure. 

\noindent {\bf Comparison of U.S. and global effort}\\

The U.S. and global efforts contine strongly on the experimental and
theoretical fronts for hybrid meson studies.  Evidence from completely
different experiments are needed to prove conclusively that the previous
hybrid meson evidence is not the result of some incorrectly interpreted
artifice.

COMPASS uses a 200 GeV pion beam and photon/Pomeron targets, and magnetic
spectrometers and EM calorimeters, to measure completely pion-photon and
pion-Pomeron reactions.  In Primakoff scattering, a high energy pion
scatters from a virtual photon in the Coulomb field of the target nucleus;
while for Diffractive scattering, the pion scatters from an exchanged
Pomeron. The relevant Primakoff and Diffractive reactions are:\\ $\pi^-
\gamma ~or~ \pi^- Pomeron \rightarrow Hybrid \rightarrow \rho\pi, \eta\pi,
\eta'\pi, b_1(1235)\pi, \pi f_1$, etc.\\ Consider some typical
experimental angular distributions, such as those for the $\pi^- \gamma
\rightarrow \eta\pi$ for different values of the $\eta\pi$ invariant mass. 
If these events are associated with the production and decay of a $J^{PC}=
1^{-+}$ hybrid state, which are quantum numbers not possible for a regular
$q\bar{q}$ meson, then a detailed partial wave analysis (PWA) of a large
data sample of such $\eta\pi$ events (centered at a given mass, with a
given width, and with appropriate resonance phase motion)  would require
these quantum numbers.  In COMPASS, the relative strengths of hybrid
compared to other close lying states, may improve significantly compared
to previous production experiments.  This may diminish the PWA
uncertainties that affected previous hybrid production experiments \cite
{hybrid}.

For Primakoff scattering, the hybrid meson production cross section
depends on the strength of its $\pi\rho$ coupling, since the virtual
photon target behaves like a $\rho$ ~by vector dominance model. For
diffractive production, the relative strengths depend on the decay modes
of the tensor glueball (2$^{++}$), since the Pomeron is thought to be on
the Regge trajectory corresponding to the tensor glueball with a presumed
mass around 2 GeV. Corresponding to the glueball decay $G(2^{++}) 
\rightarrow \pi^+ Hybrid$, one expects diffractive production via $\pi^-
G(2^{++}) \rightarrow Hybrid$.  This is an additional advantage of the
COMPASS hybrid meson study \cite {hybrid}. COMPASS can look forward to two
complementary hybrid production modes, increasing our chance for achieving
a decisive advance on our understanding of the meson constituents. 
Furthermore, in Primakoff (photon exchange) and diffractive (glue-rich
pomeron exchange)  hybrid production, meson exchange backgrounds are
strongly suppressed, which is an important advantage compared to previous
Hybrid production experiments.

For the short term, the COMPASS experiment should be in production
for the muon physics program, including setup of initial Primakoff physics
programs. Monte Carlo simulations for the hybrid studies can be completed.
Further VES and CB data (from analysis of past experiments) and
theoretical calculations will become available. For the long term, COMPASS
and GSI and JLab may run and analyze and compare data for hybrid meson
structure.

\noindent {\bf Other issues}

COMPASS has a 2 meter diameter EM ECAL2 calorimeter, which is so far
instrumented with ADC readout for the central 1 meter diameter,
appropriate to pion polarizability studies.  For the hybrid meson program,
a modest COMPASS upgrade will be required to achieve ADC readout
electronics (via individual or multiplexing readout) for the full 2 meter
diameter calorimeter coverage. This would allow detection of $\eta$s from
hybrid decay ($\eta$ or $\pi f_1$ followed by $f_1 \rightarrow
\pi\pi\eta$, etc.)  with high and flat acceptance. 

\end{document}